\documentclass[aps,12pt,eqsecnum,tightenlines,showpacs,nofootinbib,endfloats,amsmath,amssymb]{revtex4}
\usepackage{graphicx}
\usepackage{bm}
\usepackage{makeidx}

\makeindex
\usepackage{graphicx}
\usepackage{bm}
\usepackage{makeidx}

\makeindex
\begin{document}
\topmargin -2cm
 \author{Alberto A. Garcia--Diaz}
 \altaffiliation{aagarcia@fis.cinvestav.mx}
 \affiliation{Departamento~de~F\'{\i}sica,
 ~Centro~de~Investigaci\'on~y~de~Estudios~Avanzados~del~IPN,\\
 Apdo. Postal 14-740, 07000 M\'exico DF, M\'exico,\\
 and\\
Departamento de F\'{\i}sica, Universidad Aut\'onoma Metropolitana--Iztapalapa,\\
A.P. 55--534, M\'exico D.F. 09340, M\'exico.\\}

 \title{Hydrodynamic equilibrium of a static star in the presence of a cosmological
constant in $2+1$--dimensions}
 \date{\today}

\begin{abstract}
Under the hydrodynamic equilibrium Buchdahl's conditions on the
behavior of the density and the pressure, for regular fluid static
circularly symmetric star in $(2+1)$--dimensions in the presence of
a cosmological constant, is established that there are no bounds
from below on the pressure and also on the mass, except for their
positiveness. The metric for a constant density distribution is
derived and its matching with the external static solution with a
negative cosmological constant is accomplished. Some mistakes of
previous works on the topic are pointed out.
\vspace{0.5cm}\pacs{04.20.Jb, 04.50.+h}
\end{abstract}

\maketitle \tableofcontents
\section{ Introduction}
This work is devoted to the study of the equilibrium for static
perfect fluid solutions with cosmological constant
generalizing in this manner the Cruz--Zanelli~\cite{Cruz:1994ar}  analysis.\\
The main objective of this contribution is to establish that in
$2+1$--dimensions there is no room for bounds of the mass
distribution.\\
{\bf Theorem}\\
 If a perfect fluid distribution fulfills the
conditions:
\begin{itemize}
\item it is described by a one-parameter state equation $p=p(\mu)$ ,
\item the density is positive, $\mu>0$, and monotonically decreasing, $\frac {d\mu}{dr}<0$,
\item it is microscopically stable, $\frac {dp}{d\mu}\geq0\rightarrow \frac{dp}{dr}\leq0,$
\end{itemize}
then, in $2+1$--dimensions, there is {\it not} a bound on the mass to the radius ratio.\\

\section{Static circularly symmetric
perfect fluid $2+1$--solution with $\Lambda$}

This section is devoted to the search of solution to the Einstein's
equations for a static $2+1$ metric in curvature coordinates
\begin{eqnarray}\label{CZ0}
&&g=-{\rm e}^{2\,\nu(r)}dt^2+{\rm
e}^{2\,\lambda(r)}\,dr^2+r^2d\theta^2.
\end{eqnarray}
for a perfect fluid\label{perfectfluid} in the presence of a
cosmological constant. The fluid is described by the
energy--momentum tensor $T_{\mu\nu}=(p(r)+\mu(r))u_\mu u_\nu+
p(r)\,g_{\mu\nu},$$\,u_{\alpha}={\rm
e}^{\nu(r)}\delta_{\alpha}^t$.\\

The Einstein equations for a perfect fluid with $\Lambda$ can be
given as
\begin{eqnarray}
&&{E^\beta}_{\alpha}:={R^\beta}_{\alpha}-\frac{1}{2}{\delta^\beta}_{\alpha}R
-\kappa{T^\beta}_{\alpha}+\Lambda{\delta^\beta}_{\alpha},\nonumber\\
&& {T^\beta}_{\alpha}=-\rho(r){\delta^\beta}_{t}{\delta^t}_{\alpha}
+p(r)\,{\delta^\beta}_{r}{\delta^r}_{\alpha}+p(r)
\,{\delta^\beta}_{\theta}{\delta^\theta}_{\alpha}.
\end{eqnarray}
Explicitly, for the metric (\ref{CZ0}), one obtains
\begin{subequations}\label{cruzEins}
\begin{eqnarray}\label{cruzEins1}
&& E^1_1=0:{\frac {d}{dr}}\lambda  =\,r\,(\kappa\,\mu+\Lambda){ {\rm
e}^{2\,\lambda  }}, \rightarrow{\rm e}^{-2\,\lambda \left( r \right)
}=C_0-2\int^r(\kappa\mu+\Lambda
)\,r\,dr=C_0-\frac{\kappa}{\pi}\,m\left( r \right), \nonumber\\&&
m\left( r \right):=2\pi\int^r\,r\,(\mu+\Lambda/\kappa)
\,dr=\int^r\,2\pi\,r\,\mu\,dr+\frac{\pi}{\kappa}\Lambda\,r^2
=:M(r)+\frac{\pi}{\kappa}\Lambda\,r^2,
\end{eqnarray}
\begin{eqnarray}\label{cruzEins2}
E^2_2=0:EQ_{\nu1}:={\frac {d\nu}{dr}}  -r{{\rm e}^{2\,\lambda \left(
r
 \right) }} \left( \kappa\,p  -\Lambda \right)=0\,\rightarrow \nu
 =\int^r \frac{{\left( \kappa p -\Lambda \right)\,r\,dr}}
 {C_0-\frac{\kappa}{\pi}\,m\left( r \right)},
\end{eqnarray}
\begin{eqnarray}\label{cruzEins3}
E^3_3=0:EQ_{\nu\,2}:={\frac {d^{2}}{d{r}^{2}}}\nu  + \left( {\frac
{d}{dr}} \nu  \right) ^{2}- \left( {\frac {d}{dr}}\nu \right) {\frac
{d}{dr}}\lambda
 +(\Lambda-\kappa\,p
  )\,{
{\rm e}^{2\,\lambda  }}=0.
\end{eqnarray}
\end{subequations}

The substitution of the derivative ${\frac {d\nu}{dr}}$  from
$(\ref{cruzEins2})$ into $(\ref{cruzEins3})$ yields the same
equation arising from the energy--momentum conservation law
${T^{\alpha\beta}}{;\beta}=0$, namely the
Tolman--Oppenheimer--Volkoff equation:
\begin{eqnarray}\label{cruzEinsC4}
{\frac {dp}{dr}}  =- \, \left( \mu  +p
 \right)\, {\frac {d\nu}{dr}}=-{{\rm e}^{2\,\lambda }}r \left( \kappa\,p  -\Lambda \right)
 \left( \mu  +p \right)
\rightarrow{\frac {d\,p }{dr}}=-{\frac {\,r \left( \kappa\,p
 -\Lambda \right)  \left( \mu +p
 \right) }{{ C_0}-\kappa\,M \left( r \right)/\pi
-\Lambda\,{r}^{2}}}.
\end{eqnarray}
An integral quantity arises from the combination of
(\ref{cruzEins2}) and  (\ref{cruzEins3}),
$(EQ_{\nu\,2}-EQ_{\nu\,1}/r){\rm e}^{\nu(r)}$, namely
\begin{eqnarray}
\frac{d}{dr}\left( \frac{{\rm e}^{-\lambda }}{r}{\frac {d{\rm
e}^{\nu }}{dr}}\right) =0,
\end{eqnarray}
 On the
other hand, the substitution of $p(r)$ from $(\ref{cruzEins2})$ into
$(\ref{cruzEins3})$  gives rise to
\begin{eqnarray}
{\frac {d^{2}\nu}{d{r}^{2}}}  + \left( {\frac {d\nu}{dr}}
  \right) ^{2}- \left( {\frac {d\nu}{dr}}  \right) {\frac {d\lambda }{dr}} -{\frac {1 }{r}}{
\frac {d\nu}{dr}} =0,
\end{eqnarray}
which is a first order equation for $N(r):={\frac {d\nu}{dr}}$,
which can be written in a very simple form by introducing the
functions
\begin{eqnarray}\label{nueq1}
 \xi(r):=r\,{\rm e}^{\lambda(r)},\, Z:={N}/{\xi},
\end{eqnarray}
namely
\begin{eqnarray}\label{nueq2}
\frac{d}{dr}\left(\frac{N}{\xi}\right)+(\frac{N}{\xi})^2\xi=0\rightarrow
{d Z^{-1}=\xi\,dr}\rightarrow Z^{-1}=C_1+\int^r_0\xi(r)dr.
\end{eqnarray}
The equation for $\nu$ becomes
\begin{eqnarray}
&&{d{\nu}}=Z\xi\,{dr}=
\frac{\xi\,{dr}}{C_1+\int^r_0\xi(r)dr}\rightarrow
{\nu}(r)=\ln{[C_1+\int^r_0\xi(r)dr]}+\ln{C_2/2},\nonumber\\&& \nu
\left( r \right) =\ln  \left( \int^r_0 \!{{\rm e}^{\lambda \left( r
 \right) }}r{dr}+{C_1} \right) +\ln{C_2/2},
\end{eqnarray}
the const $C_2/2\rightarrow 1$ by scaling the time coordinate.\\
Substituting this integral in $(\ref{cruzEins2})$ one obtains the
pressure
\begin{eqnarray}
p \left( r \right) ={\frac {1}{\kappa\,{{\rm e}^{\lambda \left( r
 \right) }} \left( \int^r_0 \!{{\rm e}^{\lambda \left( r \right) }}r{dr}+{
C_1} \right) }}+{\frac {\Lambda}{\kappa}}
\end{eqnarray}
Summarizing, the structural functions can be given as
\begin{eqnarray}
&&{\rm e}^{\lambda \left( r \right)} =1/\sqrt{ {C_0}-\frac
{\kappa}{\pi}\,\int^r_0(2\pi\,\mu(\tilde r)\,\tilde r\,d\tilde r)
 -\Lambda\,{r}^{2}},
 \nonumber\\&&
{\rm e}^{\nu \left( r \right)} =C_1+\int^r_0[\, r^{'}\,d
r^{'}/\sqrt{ {C_0}-\frac
{\kappa}{\pi}\,\int^{r^{'}}_0(2\pi\,\mu(\tilde r)\,\tilde r\,d\tilde
r)
 -\Lambda\,{r}^{2}}].
\end{eqnarray}
 Finally, the pressure results in
\begin{eqnarray}\label{pressEXP}
\kappa\,p \left( r \right) =\frac{\sqrt{ {C_0}-\frac
{\kappa}{\pi}\,\int^r_0(2\pi\,\mu(\tilde r)\,\tilde r\,d\tilde r)
 -\Lambda\,{r}^{2}}}{C_1+\int^r_0[\, r^{'}\,d r^{'}/\sqrt{ {C_0}
 -\frac {\kappa}{\pi}\,\int^{r^{'}}_0(2\pi\,\mu(\tilde r)\,\tilde r\,d\tilde r)
 -\Lambda\,{r}^{2}}]}
 +\Lambda.
 \end{eqnarray}
This relation determines the pressure $p$ through the energy density
$\mu$ in a functional manner, if $p$ were expressed by a state
equation of the form $p=p(\mu)$, the equation (\ref{pressEXP}) gives
rise to an integral differential equation for the energy as function
of the variable $r$.

The pressure has to vanish at boundary circumference $r_b$ of the
circle, $p(r_b)=0$, where the mass function $M(r)$ determines the
total mass of the fluid $M(r_b)$ on the circle. Because the metric
signature has to be preserved throughout the whole space--time, the
positiveness of $g_{rr}$ imposes an upper bound on the value of the
total mass, namely
\begin{eqnarray}\label{cruzEinsMass}
M(r_b)\leq\frac{\pi}{\kappa}(C_0-\Lambda\,r_b^2).
\end{eqnarray}

\subsection{Cotton tensor types}

The Cotton tensor for this perfect fluid occurs to be
\begin{eqnarray}\label{cruzEinsCotton}
{C^\mu}_\nu=-\frac{\kappa}{4}\, {\rm e}^{-\lambda}\,{ \frac {d
\mu(r)}{dr}} (r\,{\rm
e}^{-\nu}\delta^\mu_t\delta_\nu^\theta-\frac{{\rm
e}^{\nu}}{r}\delta^\mu_\theta\delta^t_\nu).
\end{eqnarray}
Therefore, the solution with constant density $\mu_0$ is conformally
flat, vanishing Cotton tensor.\\
Moreover, the search for its eigenvectors yields
\begin{eqnarray}
\lambda_1=0;{\bf V}1&&=[0,V^{2},0],\,V1_{\mu}=
V^{2}g_{rr}\delta^r_{\mu},\,V1^{\mu}V1_{\mu}=
(V^{2})^2g_{rr},\,{\bf V}1={\bf S}1,\nonumber\\
\lambda_{2}=i\frac{\kappa}{4}{\rm e}^{-\lambda}\,{ \frac {d
\mu(r)}{dr}} ;{\bf V}2&&=[V^{1},0, V^{3}=-{\frac {i{{\rm e}^{\nu }}{
V^{1}}}{r}}],\,\,{\bf V}2={\bf Z},
\nonumber\\
\lambda_{3}=-i\frac{\kappa}{4}{\rm e}^{-\lambda}\,{ \frac {d
\mu(r)}{dr}} ;{\bf V}3&&=[V^{1},0, V^{3}={\frac {i{{\rm e}^{\nu }}{
V^{1}}}{r}}],\,\, {\bf V}3={\bf {\bar Z}},
\end{eqnarray}
consequently the corresponding tensor type is
$${\rm Type:}\{S,Z,\bar Z\}.$$
\section{Cruz--Zanelli existence of hydrostatic equilibrium for
$\Lambda\leq0$}

In the work~\cite{Cruz:1994ar} it is established that: A perfect
fluid in hydrostatic equilibrium,
 $$(\mu (r\leq r_b)> 0 , \,p(r\leq r_b)>0,\,
\frac {dp}{dr}|_{r\leq r_b} \,\,\text{pressure monotonically decreasing}),$$
is only possible for $\Lambda\leq0$. \\
The condition on $\Lambda$ follows from the energy--momentum
conservation equation (\ref{cruzEinsC4})
$$
{\frac {d}{dr}}p \left( r \right) =-{\frac {r\, \left( \kappa\,p
  -\Lambda \right)  \left( \mu  +p \right) }{ C_0-\kappa\,M \left( r \right)/\pi
-\Lambda\,{r}^{2}}},
$$  \\
which evaluated at the boundary yields
$${\frac {dp}{dr}}|_{r=r_b} =\frac{\,r_b \Lambda }
{{C_0}-\Lambda\,r_b^2-{\kappa}\,M\left( r_b \right)/\pi} \mu ( r_b), $$
which for $\mu ( r)\geq0 $, and $M(r_b)$ fulfilling
(\ref{cruzEinsMass}) is non--positive only if $\Lambda\leq0.$
\\Moreover, since for $\Lambda\leq0$ the right--hand side of
$$\frac {d}{dr}p \left( r \right) =-{\,r(\kappa\,p- \Lambda )}
\left(p  +\mu \right)g_{rr}$$
is always negative, then $p \left( r \right)$ is a decreasing
function such that $$p \left( r=0 \right)=p_c> p\left( r_b
\right)=0.$$

\section{ Positiveness of the hydrostatic pressure}

Form (\ref{cruzEins1}) one has
$$\frac {dM}{dr}=2\pi\,r\,\mu(r)$$
which combined with (\ref{cruzEins3}) gives
\begin{eqnarray}\label{cruzEinsMassN}
\frac {d}{dr}\left(p-A\,M(r)
\right)=-(\kappa\,p+\frac{1}{l^2})(p+\mu(r))g_{rr}-2\,A\pi\,r\,\mu(r)
\end{eqnarray}
since the right--hand side of this equation, for positive $A=\rm
const>0$, is always negative, then

$${\textsc{R}(r) }=p-A\,M(r)$$
is monotonically decreasing, and ${\textsc{R}(0) }>{\textsc{R}(r_b)
}\rightarrow p_c-A\,M(0)>-A\,M$, and consequently
$p_c>-A\,(M-M(0))$, which means that the hydrostatic pressure cannot
be negative.

Since the right--hand side of (\ref{cruzEinsMassN}) for $A=0$ is
always negative, then $p \left( r \right)$ is a monotonically
decreasing function, and consequently $p \left( r=0 \right)=p_c>
p\left( r_b \right)=0\rightarrow p_c>0.$

In no way for the choice of $A=-(2\,\pi\,l^2\,C_0)^{-1}<0$,
equivalent to the one done in CZ~\cite{Cruz:1994ar}\,(13), one could
establish the bound ~\cite{Cruz:1994ar}CZ\,(10).

\section{Buchdahl's theorem in $2+1$ hydrostatics}

If a perfect fluid distribution fulfills the conditions:
\begin{enumerate}
\item it is described by a one-parameter state equation $p=p(\mu)$ ,
\item the density is positive $\mu>0$ and monotonically decreasing $\frac {d\mu}{dr}<0$,
\item it is microscopically stable, $\frac {dp}{d\mu}\geq0$, $(\frac{dp}{dr}\leq0),$
\end{enumerate}
then there is not a bound on the density.

The Lemma on no existence of a bound for the density is based on
Einstein equations (\ref{cruzEins2}) and  (\ref{cruzEins3}), which
yield
\begin{eqnarray}\label{cruzEinsEQ1}
(EQ_{\nu\,2}-EQ_{\nu\,1}/r){\rm
e}^{\nu(r)}=\frac{r}{\sqrt{{{C_0}}-\frac{\kappa}{\pi}\,m(r)}}\frac{d}{dr}
[\frac{1}{r}\sqrt{{{C_0}}-\frac{\kappa}{\pi}\,m(r)}\frac{d}{dr}{\rm
e}^{\nu(r)}]=0,
\end{eqnarray}
because $EQ_{\nu\,1}=0=EQ_{\nu\,2}$, therefore
\begin{eqnarray}\label{cruzEinsEQ2}
&& [\frac{1}{r}\sqrt{{{C_0}}-\frac{\kappa}{\pi}m(r)}\frac{d}{dr}{\rm
e}^{\nu(r)}]=C_{\nu}={\rm const.},\, d\,{\rm
e}^{\nu(r)}=C_{\nu}\frac{\sqrt{{{C_0}}-\frac{\kappa}{\pi}m(r)}}{r}\,dr
=C_{\nu}\,\xi(r)\,dr,
\nonumber\\&& \rightarrow{\rm
e}^{\nu(r)}=C_{\nu}\left(C_1+\int^r_0{r\,{\rm
e}^{\lambda(r)}}dr\right),\,(C_{\nu}\rightarrow1,\,t\rightarrow
t/\sqrt{C_{\nu}}),
\end{eqnarray}
which are no others than the equations for $\nu$ and its
differential given in (\ref{nueq2}). Thus one cannot establish a
condition on the energy density in contrast with the statement of
~\cite{Cruz:1994ar}\,CZ\,(15), which is based on a
miss--interpretation of equations ~\cite{Cruz:1994ar}\,CZ\,(A4) and
~\cite{Cruz:1994ar}\,CZ\,(A5).

The equation CZ~\cite{Cruz:1994ar}\,(A1):
\begin{eqnarray*}\label{cruzEinsEQQQ}
\frac{d}{dr}[\frac{d\nu}{dr}(\frac{2\pi
{C_0}}{\kappa}-2m(r))/r]+[\frac{d\nu}{dr}(\frac{2\pi
{C_0}}{\kappa}-2m(r))/r+\frac{d\,m}{dr}]\frac{d\nu}{dr}=0.
\end{eqnarray*}
is equivalent to (\ref{cruzEinsEQ1}), which by introducing $
\gamma=\int^r_0\,(\frac{2\pi {C_0}}{\kappa}-2m(r))^{-1/2}\,r\,d\,r $
can be written as
\begin{eqnarray*}\label{cruzEinsEQQQ}
\frac{d^2}{d\gamma^2}{\rm
e}^{\nu(r)}=0\rightarrow{\frac{d}{d\gamma}{\rm
e}^{\nu(r)}}=C_{\nu}={\rm const.},\rightarrow{\rm
e}^{\nu(r)}=C_{\nu}\gamma+{\rm e}^{\nu(0)}\neq\,{\rm
e}^{\nu(r)}=\gamma\frac{d{\rm e}^{\nu(r)}}{d\gamma}+{\rm
e}^{\nu(0)}\,{\rm (A5)}.
\end{eqnarray*}
The equation ~\cite{Cruz:1994ar}\,CZ\,(A5) is the source of the
mistake.

\section{Static star with constant density $\mu_0$
and $\Lambda=-1/l^2\leq0$ }

The static star with uniform density $\mu_0$ is characterized by
mass and pressure given respectively as:
 \begin{eqnarray}\label{cruzEinsMass2L}
&&M(r)=\pi\,\mu_0\,r^2,m(r)=\pi\,\mu_0\,r^2+\frac{\pi}
{\kappa}\Lambda\,r^2\nonumber\\
&&
p(r)=\mu_0\frac{\sqrt{{C_0}-\frac{\kappa}{\pi}m(r)}-\sqrt{{C_0}
-\frac{\kappa}{\pi}m(r_b)}}
{-\sqrt{{C_0}-\frac{\kappa}{\pi}m(r)}
-\frac{\kappa}{\Lambda}\mu_0\,\sqrt{{C_0}-\frac{\kappa}{\pi}m(r_b)}}.
\end{eqnarray}
A star of uniform density in hydrostatic equilibrium
(${\Lambda}=-1/l^2$) possesses central mass and pressure of the form
\begin{eqnarray}\label{cruzEinsMass2L}
&&M(0)=0,m(0)=0,\nonumber\\
&&
p_c=\mu_0\frac{\sqrt{{C_0}}-\sqrt{{C_0}-\frac{\kappa}{\pi}m(r_b)}}
{-\sqrt{{C_0}}+{\kappa\,l^2}\mu_0\,\sqrt{{C_0}-\frac{\kappa}{\pi}m(r_b)}},
\end{eqnarray}
at the boundary $M(r_b)=\pi\,r_b^2\mu_0=\mu_0\,S_{\bigodot},
p(r_b)=0$, while for
$r_b=\sqrt{\frac{{C_0}}{\kappa\,\mu_0}}\sqrt{1+\frac{1}{\kappa\,l^2\,\mu_0^2}}$
the pressure becomes infinity, $p\rightarrow\,\infty,$ and the mass
equates to
$M=\frac{\kappa\,{C_0}}{\pi}(1+\frac{1}{\kappa\,l^2\,\mu_0})$,
$(m=\frac{\kappa\,{C_0}}{\pi}(1-\frac{1}{\kappa^2\,l^4\,\mu_0^2}))$.

The evaluation of ${\rm e}^{2\nu}$  or the metric component
$g_{tt}=-{\rm e}^{2\nu}$  yields
\begin{eqnarray}
{\rm
e}^{2\nu}=\left(\frac{\kappa\,\mu_0\,l^2
\sqrt{{C_0}-\frac{\kappa}{\pi}m(r_b)}-\sqrt{{C_0}-\frac{\kappa}{\pi}m(r)}}
{\kappa\,\mu_0\,l^2-1}\right)^2,
\end{eqnarray}
which is a slightly different expression compared with the one of
CZ~\cite{Cruz:1994ar}(23) which is given with  an extra
multiplicative factor
$({r^2/l^2-M_0})/{\sqrt{{C_0}-\frac{\kappa}{\pi}m(r_b)}}$.

The external solution to which the uniform fluid solution can be
matched is the static anti--de Sitter metric with parameter $M_0$
known also as the static BTZ solution
\begin{eqnarray}
g=-(-M _0+\frac{r^2}{l^2})dt^2+dr^2/(-M _0+\frac{r^2}{l^2})+r^2
d\phi^2
\end{eqnarray}
the continuity at the boundary $r_b$ of the metric for the fluid is
achieved
\begin{eqnarray}
{\rm e}^{2\nu(r_b)}={\rm e}^{-2\lambda(r_b)}={C_0}+\frac{r_b^2}{l^2}
-\frac{\kappa}{\pi}M=-M
_0+\frac{r_b^2}{l^2}\rightarrow\,{\kappa}={\pi},\,M(r_b)=M=M
_0-{C_0}.
\end{eqnarray}
The static perfect fluid solution with $\Lambda=-1/l^2$ could
exhibit an event horizon at
$r_h=\frac{C\,l^2}{\kappa\,\mu_0\,l^2-1}.$

\section{Acknowledgments}
This work has been carried out during a sabbatical year at the
Physics Department--UAMI, and has been partially supported by Grant
CONACyT 178346.

\end{document}